%% file: retrobhb.tex
\documentclass[fleqn,usenatbib]{mnras}

\usepackage{newtxtext,newtxmath}
\usepackage[normalem]{ulem} %For sout, can comment later
\usepackage[T1]{fontenc}

\DeclareRobustCommand{\VAN}[3]{#2}
\let\VANthebibliography\thebibliography
\def\thebibliography{\DeclareRobustCommand{\VAN}[3]{##3}\VANthebibliography}

\usepackage{graphicx}	% Including figure files
\usepackage{subfig}
\usepackage{amsmath}	% Advanced maths commands
\usepackage{tikz}
\usepackage[T1]{fontenc}
\usepackage{aecompl}
\usepackage{calligra}
\DeclareMathAlphabet{\mathcalligra}{T1}{calligra}{m}{n}
\DeclareFontShape{T1}{calligra}{m}{n}{<->s*[2.2]callig15}{}

\usepackage{hyperref}
\hypersetup{
    pdfnewwindow=true,      % links in new window
    colorlinks=true,       % false: boxed links; true: colored links
    linkcolor=orange,      % color of internal links
    citecolor=cyan,        % color of links to bibliography
    filecolor=orange,      % color of file links
    urlcolor=orange           % color of external links
}
\usepackage{units}
\usepackage{color}

\def\MSun{{{\rm M}_{\odot}}}

\def\fedd{{\rm f}_{\rm Edd}}
\def\chiyr{\chi_{{\rm eff},1{\rm yr}}}

\def\eg{{\em e.g.}}
\def\ie{{\em i.e.}}

\title[$\chi_{\rm eff}$ vs $e$ for MBHBs in gas]{Accretion mediated spin-eccentricity correlations in LISA massive black hole binaries}
\author[M. Garg et al.]{Authors et al.}

\author[Garg, Tiede $\&$ D'Orazio]{
Mudit Garg,$^{1}$\thanks{E-mail: mudit.garg@uzh.ch}
Christopher Tiede,$^{2}$ \& 
Daniel J. D'Orazio$^{2,3}$
\\
$^{1}$Department of Astrophysics, University of Zurich, Winterthurerstrasse 190, CH-8057 Z\"urich, Switzerland \\
$^{2}$Niels Bohr International Academy, Niels Bohr Institute, Blegdamsvej 17, 2100 Copenhagen, Denmark\\
$^{3}$Space Telescope Science Institute, 3700 San Martin Drive, Baltimore, MD 21218, USA
}

\date{Accepted ???. Received ???; in original form ???}

\pubyear{2024}

\begin{document}
\label{firstpage}
\pagerange{\pageref{firstpage}--\pageref{lastpage}}
\maketitle

% Abstract of the paper
\begin{abstract}
\noindent
We examine expected effective spin ($\chiyr$) and orbital eccentricity ($e_{1\rm yr}$) correlations for a population of observable equal-mass massive black hole binaries (MBHBs) with total redshifted mass $M_z\sim[10^{4.5},10^{7.5}]~\MSun$ embedded in a circumbinary disc (CBD) at redshifts $z=1$ and $z=2$, one-year before merging in the LISA band. We find a strong correlation between measurable eccentricity and negative effective spin for MBHBs that are carried to merger by retrograde accretion. This is due to the well-established eccentricity pumping of retrograde accretion and less-well-established formation of retrograde minidiscs coupled with a stable retrograde CBD throughout the binary evolution from the self-gravitating radius. Conversely, prograde accretion channels result in positive $\chi_{{\rm eff},1\rm yr}$ and non-measurable $e_{1\rm yr}$ except for almost unity Eddington ratio and $M_z\lesssim10^{5}~\MSun$ MBHBs at $z=1$. This clear contrast between the two CBD orientations -- and particularly the unique signature of retrograde configurations -- provides a promising way to unlock the mysteries of MBHB formation channels in the LISA era.
\end{abstract}

%

%%%%%%%%%%%%%%%%% BODY OF PAPER %%%%%%%%%%%%%%%%%%

% =============================================================================
\section{Introduction} 
\label{S:Introduction}

Galaxy mergers in the Universe lead to the formation of massive black hole binaries (MBHBs; \citealt{Begelman1980}). Dynamical friction from stars and gas can bring such MBHBs to parsec scales, but beyond this it becomes inefficient. At these separations, gravitational waves (GWs) alone can not merge the binary within the lifetime of the Universe. This is canonically referred to as the final-parsec problem \citep{Milosavljevic2002}. However, there are several possibilities suggested to overcome this problem such as improvements in modeling stellar potentials \citep[see, e.g.][]{Preto2011, Khan2011,Vasiliev2015}, subsequent galactic mergers, and gas discs around binaries \citep[see, \eg,][]{AmaroSeoane2022}, but how this is overcome in nature remains uncertain. These dynamical effects can shrink MBHBs through the sub-parsec regime until GWs become strong enough to drive the binary to merger within a Hubble time. Recently, pulsar timing arrays \citep{EPTAInPTA2023,Agazie2023,Reardon2023,Xu2023} have provided evidence for a stochastic GW background that is likely sourced by a population of $\gtrsim 10^8\MSun$ inspiraling MBHBs, which have made it through the final parsec. If this also occurs for $10^4-10^8~\MSun$ near-equal-mass merging MBHBs, then they will be observable up to redshift $z\sim20$ by the Laser Interferometer Space Antenna (LISA; \citealt{AmaroSeoane2017,Colpi2024}), recently adopted by ESA, together with in-development TianQin \citep{Li2024} and Taiji \citep{Gong2021}.

In the LISA milli-Hz-frequency band, GWs from MBHBs will have high signal-to-noise ratios (SNRs), which will help to constrain source parameters and also put constraints on  merging binaries' environments. The effects of gas are particularly compelling because they can simultaneously provide electromagnetic (EM) counterparts \citep{Haiman2017,Mangiagli2022, DOrazioCharisi+2023,Franchini2024,Cocchiararo2024} and leave detectable imprints on the GW waveform in LISA. For the latter, gas has been demonstrated to alter the binary inspiral rate \citep{Garg2022,DOrazio2021,Garg2024b,Tiede2024b}, to produce measurable residual orbital eccentricities \citep{RoedigSG_retro:2014,Zrake2021,Tiede2024b,Fumagalli2023,Garg2024a}, and to impact the magnitude and orientation of the component BH spins \citep{Bardeen1970,Thorne1974,Bardeen1975}. Similar to gas, stellar hardening \citep{Quinlan1996,Khan2011,Gualandris2022} and three-body interactions \citep{Blaes2002,Bonetti2019} can produce measurable residual binary eccentricities, but only the latter can additionally alter the component BH spin orientation \citep{Liu2017}. Lastly, ignoring even mild gas effects or small orbital eccentricity when modeling MBHBs in the LISA band can suggest false violations of general relativity \citep{Garg2024d}.

In the classical picture for massive binary accretion, radiatively efficient gas settles into a co-planar circumbinary disc (CBD; \citealt{DOrazio2016}) with prograde (retrograde) CBDs leading to CBD-aligned mini-discs around each component BH \citep[e.g.][]{Farris2014, Tiede2024b}. Recent suites of high-resolution hydrodynamical simulations of CBDs suggest pumping or damping of binary orbital eccentricity depending on binary and disc parameters. In particular, whether the disc settles into a prograde or retrograde configuration can have an important impact on the binary eccentricity evolution, and the rate at which the binary inspirals (if at all; see, \eg, \citealt{Munoz:2019}) which could lead to distinguishable parameters in the LISA band for the same starting binary. For prograde accretion scenarios \citep{Zrake2021,DOrazio2021,Siwek2023}, gas brings the binary to an equilibrium eccentricity $\sim0.4$, where it shrinks the binary's semi-major axis until GWs are strong enough to drive the binary to coalescence, circularizing the orbit \citep{Peters1963,Peters1964}. In retrograde configurations \citep{BankertShiKrolik:2015, Schnittman_Krolik_retro:2015, Tiede2024b}, there is an eccentricity pumping throughout the evolution until the GWs takeover (but, see also \citealt{NixonKing_retro+2011, RoedigSG_retro:2014} for low-eccentricity behavior). In retrograde discs, the binary also shrinks its semi-major axis at least two-times more quickly than in prograde scenarios leading to the possibility of different final masses, spins of the component BHs \citep{Dotti2010,Miller2013}, and orbital eccentricity \citep{Roedig2011,Schnittman_Krolik_retro:2015,Zrake2021} near-coalescence for the same starting binary. 

A recent study \citep{Garg2024b} put constraints on the CBD properties using GWs from individual events alone. Degeneracies between disc parameters can be possibly broken by associated EM counterpart observations. Another possibility is to study the population of observed events. Realization of MBHB merger rates at the high end of what is expected for LISA (100's per year; \citealt{AmaroSeoane2022}) would allow us to not only analyze events individually, but also the population and learn underlying properties of source distributions and their environments. This strategy is already employed by ground-based GW detectors to disentangle formation channels for the observed stellar-mass binary BH population \citep{Nishizawa2016,Breivik2016,Samsing2018,DOrazio2018,RomeroShaw2019,RomeroShaw2020,Zevin2021,RomeroShaw2022}. 

Here we conduct a quantitative study of the differences between merging MBHBs in the LISA band when driven towards merger via retrograde vs prograde accretion. We show that retrograde configurations yield negative effective spin parameters ($\chi_{\rm eff}$; \citealt{Damour2001}) correlated with residual eccentricity near-merger, where both can be readily measured via GW observations. This contrasts with prograde systems and may be unique among formation channels, offering a possibly distinct indicator of the environmental effects that abetted a binary merger.

\section{Binary Evolution Model}
\label{S:Model}
\begin{figure}
\centering
\includegraphics[width=\columnwidth]{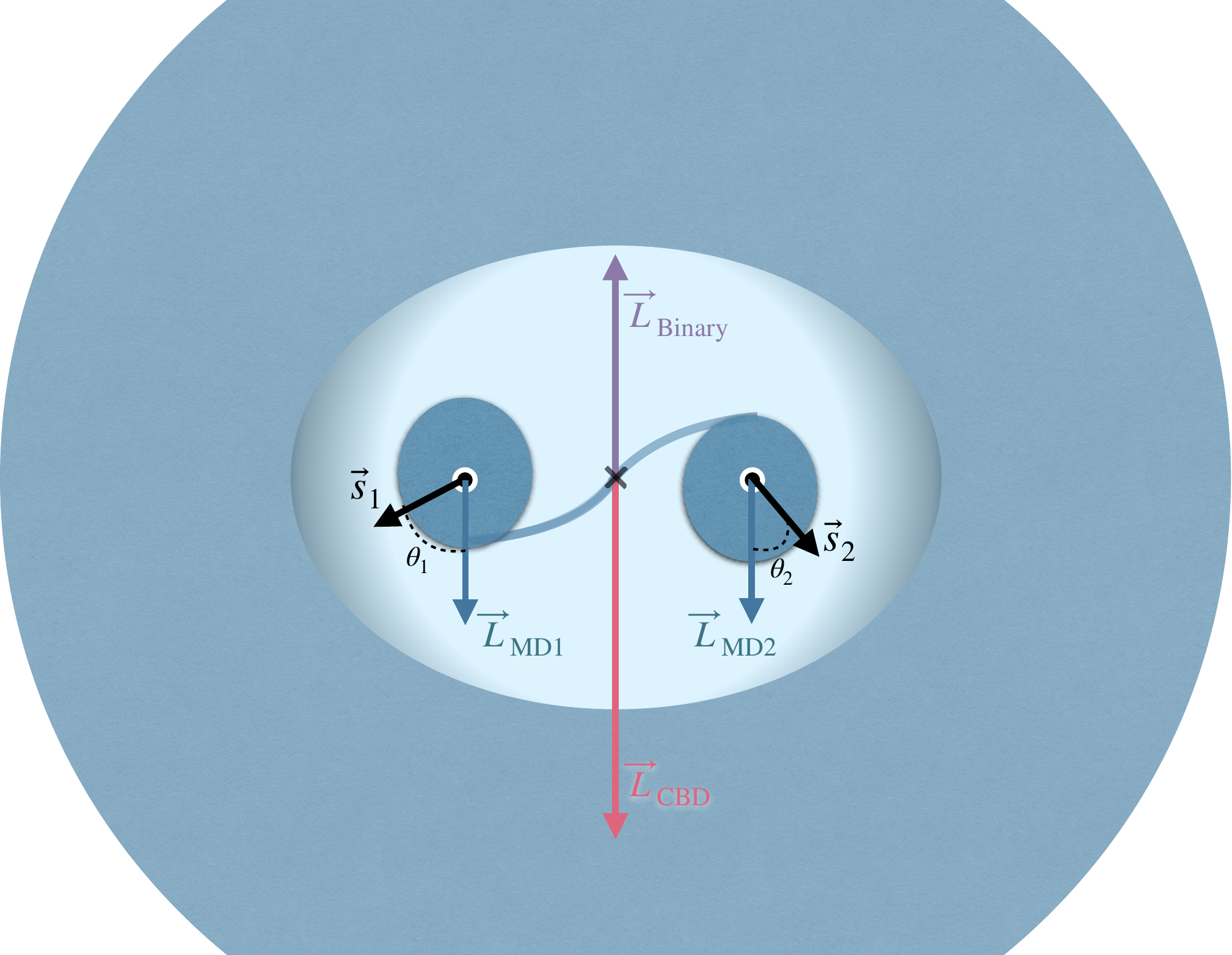}
    \caption{
    Schematic for a retrograde CBD configuration, where the binary with angular momentum $\vec{L}_{\rm Binary}$ is anti-aligned with that of the accretion disc $\vec{L}_{\rm CBD}$. The two component BHs have spin vectors $\vec{s}_{1,2}$, which are misaligned by an angle $\theta_{1,2}\in[0,\pi/2)$ with respect to their own CBD-aligned mini-discs of angular momenta $\vec{L}_{{\rm MD}1,2}$.}
    \label{fig:Doodle}
\end{figure}

Consider the motion of two equal-mass spinning BHs of total redshifted mass $M_z$ at redshift $z$ in a bound eccentric orbit with a detector-frame semi-major axis\footnote{Detector-frame SMA is $(1+z)$ times the source-frame SMA, since redshifted orbital frequency $\propto\sqrt{M_z/a^3}$.} (SMA) $a$ and eccentricity $e$ embedded in a circumbinary disc (CBD). A prograde (retrograde) CBD implies that the disc's angular momentum ($\vec{L}_{\rm CBD}$) is in the same (opposite) direction as the binary's angular momentum ($\vec{L}_{\rm Binary}$). Moreover, a prograde (retrograde) CBD leads to CBD-aligned mini-discs around each component BH \citep{DOrazio2021,Tiede2024b}. We also define a dimensionless spin parameter\footnote{Here $G$ is the gravitational constant and $c$ is the speed of light.} $s \equiv cJ/Gm^2\in(-1,1)$ for a BH of mass $m$ and a spin angular momentum $J$ with positive (negative) values corresponding to a prograde (retrograde) mini-disc orientation with respect to the BH. Fig.~\ref{fig:Doodle} shows a schematic for our system for a retrograde CBD configuration.

The quantity $s$ sets the innermost stable circular orbit (ISCO; $a_{\rm ISCO}$) and radiative efficiency ($\epsilon$) of the accretion onto the BH \citep{Fiacconi2018}:
\begin{align}
    a_{\rm ISCO}&=F(s){Gm}/{c^2},\quad
    \epsilon=1-\sqrt{1-{2}/{3F}},\\
    F(s)&=3+Z_2\mp\sqrt{(3-Z_1)(3+Z_1+2Z_2)}\nonumber,\\
    Z_1&=1+(1-s^2)^{\frac13}[(1+s)^{\frac13}+(1-s)^{\frac13}],\quad Z_2=\sqrt{3s^2+Z_1^2},\nonumber
\end{align}
where upper (lower) sign represents a prograde (retrograde) mini-disc configuration. 

The evolution of a spin ($s_i$) that is fully (anti-) aligned with its co-planar mini-disc (\ie, $\theta_i=0$) is \citep{Fiacconi2018}
\begin{align}\label{eq:chidot}
    \dot{s}_i&=\frac{\dot{M}_{z,i}}{M_{z,i}}\left(\frac{{F}^2\mp2s_i\sqrt{F}+s_i^2}{{F}\sqrt{{F}-3\pm2s_i/\sqrt{F}}}\frac{1}{1-\epsilon_i}-2s_i\right)=\frac{\dot{M}_{z,i}}{M_{z,i}}{\tilde F}(s_i),
\end{align}
where again upper (lower) sign represents a prograde (retrograde) mini-disc orientation and $M_{z,i}=M_z/2$. Here $\dot{M}_{z,i}$ is the accretion rate onto the component BH and determined by the steady-state accretion rate at infinity $\dot M_0\equiv\fedd M_{z}/45 ~{\rm Myr}^{-1}$ \citep{Jiang2014} via $\dot{M}_{z,i}=(1-\epsilon_i)\dot M_0/2$. Assuming $\dot M_0$ is split evenly between both component BHs, motivated by measured preferential accretion for both prograde and retrograde CBDs \citep{D'Orazio2024}, we can write down the accretion rate onto the binary
\begin{align}
    \dot{M}_z=\left(1-\frac{\epsilon_1+\epsilon_2}{2}\right)\dot M_0=\left(1-\frac{\epsilon_1+\epsilon_2}{2}\right)\frac{\fedd}{45~{\rm Myr}}M_z,
\end{align}
where $\fedd$ is the Eddington ratio and $\epsilon_i$ is the efficiency of the $i$-th component BH. We assume $\dot M_0$ or equivalently $\fedd$ are constant throughout the binary evolution, \ie, the feeding rate to the disc at infinity is constant.

For an equal-mass MBHB embedded in a co-planar circumbinary disc (CBD), the details of the accretion flow and binary-disc parameters dictate if the gas torques the binary to inspiral or outspiral. For a retrograde disc, the evolution of SMA and eccentricity due to gas are \citep{Tiede2024b}
\begin{align}\label{eq:aedotgas}
    \dot{a}^{\rm ret}_{\rm gas}&\approx-10\frac{\dot{M}_z}{M_z}a,\quad\dot{e}^{\rm ret}_{\rm gas}\approx\frac{\dot{M}_z}{M_z}\begin{cases} 
        20e, & e\leq0.15, \\
        3, &e>0.15.
        \end{cases}
\end{align}
While for the prograde binary, we take $\dot{e}^{\rm pro}_{\rm gas}$ from the \citet{Zrake2021} fit and the most negative value of $\dot{a}^{\rm pro}_{\rm gas}$ from \citet{DOrazio2021} to consider the fastest merger timescale:
\begin{align}\label{eq:aedotgas2}
    \dot{a}^{\rm pro}_{\rm gas}&=-5\frac{\dot{M}_z}{M_z}a \ .
\end{align}
We also always initialize prograde binaries from their equilibrium eccentricity $e_{\rm eq}^{\rm pro}=0.45$ \citep{Zrake2021,DOrazio2021,Siwek2023}, where binaries only circularize when GW emission becomes dominant over gas effects. This will result in the fastest inspiral times for the prograde case to compare with retrograde results computed later. Slower inspiral times will only exacerbate the differences we later find between the two disc configurations.

GWs remove energy and angular momentum from a compact MBHB. At Newtonian-order, orbit-averaged evolution rates for $a$ and $e$ due to emission of GWs are \citep{Peters1963,Peters1964}
\begin{align}\label{eq:aedot}
    \dot{a}_{\rm GW}&=-\frac{16}{5}\frac{G^3}{c^5}\frac{ M_z^3}{a^3}\left(1+\frac{73}{24}e^2+\frac{37}{96}e^4\right)(1-e^2)^{-\frac72},\\
    \dot{e}_{\rm GW}&=-e\frac{76}{15}\frac{G^3}{c^5}\frac{ M_z^3}{a^4}\left(1+\frac{121}{304}e^2\right)(1-e^2)^{-\frac52}\nonumber.
\end{align}
Even though retrograde configurations in \S~\ref{S:Results} will induce high eccentricities during binary evolution, we have checked that we never reach the relativistic regime to require higher post-Newtonian (PN) terms.

We choose to begin our binary evolution from the self-gravitating radius ($R_{\rm sg}$) of a thin CBD. Beyond this, the disc can fragment, or calculations including disc self-gravity are required \citep[see, \eg][]{Franchini2021}. Furthermore, we argue that this is the valid limit for the $M_z\lesssim 10^7$ MBHBs in the LISA band. For a turbulo-viscous parameter $\alpha = 0.1$ \citep{ShakuraSunyaev1973}, the self-gravitating radius can be expressed as \citep{Perego2009,Bortolas2021}
\begin{align}
    R_{\rm sg}[r_s]=&5.32\times10^6\left({\fedd}/{0.1}\right)^{-\frac{22}{45}}\left({M_z}/{10^5\MSun}\right)^{-\frac{52}{45}},
\end{align}
where $r_s\equiv 2GM_z/c^2$ is the Schwarzschild radius of the total binary mass. For MBHBs in the LISA band, $R_{\rm sg}$ is much larger than the separation scale where GW inspiral will outpace the CBD and gas effects are no longer important ($a\lesssim10^2r_s$; see, \eg~\citealt{Dittmann2023}). Hence, for the lower-mass LISA MBHBs, we are justified in using Eqs~\eqref{eq:aedotgas} and \eqref{eq:aedotgas2}, as a gravitationally stable CBD will surround such MBHBs during a long duration of inspiral where gas effects can dominate over GW-driven evolution. Therefore, we consider earlier evolution of the MBHB via stars or gravitationally unstable gas as setting our initial conditions, which will primarily dictate final values of binary parameters in \S~\ref{S:Results}.

We assume that the binary is always fully (anti-) aligned with its CBD at least by the time the system is initialized at the self-gravitating radius. For mild initial misalignment, it takes around a few Myr for complete (anti-) alignment \citep{Martin2019}. Moreover, we assume that the magnitude and misalignment angle of the component BH's spin vector evolve independently from each other. We also always assume that the BH spins (anti-) align with their minidiscs when the binary reaches the LISA band. The spin-alignment timescale is at maximum $\sim60$ Myr for an already maximally spinning BH with almost $\pi/2$ radians misalignment for $\fedd=0.1$ and azimuthal Mach number $10$ \citep{Lodato2013}. Therefore, for almost all parameters considered, the BH spins will fully align with their mini-discs, and only for a narrow set of parameters will some residual misalignment remain near merger. This can be seen later in Fig.~\ref{fig:timescale}, where only for initial eccentricities $e_0\gtrsim0.6$, inspiral timescales are less than $60$ Myr. Our spin-alignment assumption is further supported by \citet{Bourne2023}, where they almost always find spin alignment irrespective of the disc orientation and initial eccentricity.

To study the evolution of binary parameters under the influence of both GWs and CBD, we evolve the following coupled equations from $a=R_{\rm sg}$ to $1$ year before merger\footnote{SMA, one year before merging, is calculated by integrating $\dot a_{\rm GW}$ with final separation $a_{\rm ISCO}$.} in the LISA band:
\begin{align}\label{eq:coupledevo}
    \dot{a}&=\dot{a}_{\rm GW}+\dot{a}_{\rm gas},~\dot{e}=\dot{e}_{\rm GW}+\dot{e}_{\rm gas},\\
    ~\frac{\dot{M}_z}{M_z}&=\left(1-\frac{\epsilon_1+\epsilon_2}{2}\right)\frac{\fedd}{45~{\rm Myr}},~\dot{s}_1=\frac{\dot{M}_{z,1}}{M_{z,1}}{\tilde F}(s_1),~\&~\dot{s}_2=\frac{\dot{M}_{z,2}}{M_{z,2}}{\tilde F}(s_2),\nonumber
\end{align}
where $\dot{a}_{\rm gas}$ is $\dot{a}^{\rm pro}_{\rm gas}$ ($\dot{a}^{\rm ret}_{\rm gas}$) for the prograde (retrograde) CBD.

Since we consider long-term evolution under retrograde accretion, it is important to verify that the accretion remains stably anti-aligned throughout the inspiral. If not, the CBD can tilt and eventually align in a prograde configuration \citep[see, \eg][]{RoedigSG_retro:2014}. For a stable evolution, the ratio (${\rm f}_{\rm st}$) between the disc angular momentum ($L_{\rm CBD}$) and twice the binary angular momentum (2$L_{\rm Binary}$) needs to be less than unity \citep{King2005,Nixon2011}. The binary angular momentum scales with SMA as $L_{\rm binary} \propto a^{1/2}$, and the radial dependence of the disc angular momentum $L_{\rm CBD}$ is determined by the surface density profile via the integral $\int_0^{Na} {\rm d} r\Sigma(r) r^{3/2}$, where $r=Na$ denotes the radius out to which the disc influences the binary. One can see that as long as $\Sigma(r) \propto r^\gamma$ has a power-law coefficient $\gamma > -2$, stability will always increase as the binary shrinks at fixed eccentricity. This is the case for our assumed \citet{ShakuraSunyaev1973} disc in all three regions -- inner, middle, and outer -- of the disc. We show explicitly that ${\rm f}_{\rm st}$ for the inner and middle region is always much less than unity in Appendix~\ref{App:fst}, even for high eccentricity, and discuss the outer disc here, as it presents the most stringent stability constraints.

For an equal-mass MBHB (in the source frame) around a thin disc, the stability ratio for the outer disc is\footnote{Even in the source frame, we still keep symbols $M_z$ and $a$ to avoid introducing new ones, and they represent $2$--$3$ times smaller values then respective redshifted parameters for our considered redshifts of $1$ and $2$ in \S \ref{S:Results}.}
\begin{align}\label{eq:stability}
    {\rm f}^{\rm out}_{\rm st}\sim &\frac{0.39}{\sqrt{1-e^2}}\left(\frac{\alpha}{0.1}\right)^{-\frac45}\left(\frac{\fedd}{0.1}\right)^{\frac{7}{10}}\left(\frac{M_z}{10^5\MSun}\right)^{-\frac{1}{20}}\left(\frac{N}{3}\right)^{\frac74}\\
    &\times \left(\frac{a}{5.32\times10^6\, r_s}\right)^{\frac54},\nonumber
\end{align}
where we have conservatively chosen $a$ to be the self-gravitating radius for the above reference parameters, $\alpha=0.1$, $\fedd=0.1$, and $M_z=10^5~\MSun$ (consistent with \citet{Schnittman_Krolik_retro:2015}\footnote{ Once the aspect ratio in their Eq.~(26) is converted to include radial coordinate dependence \citep{Haiman2009,Haiman2022_Err}.}). For the range of parameters we consider in this study, this ratio remains less than unity for $N\lesssim3$. This is a reasonable value for $N$ given that it is derived at a maximal binary separation, at the self-gravitating radius, and that the majority of torques in the retrograde case come from the flow at $r\lesssim 2a$ \citep{Tiede2024a}.  Hence, at binary initialization, the stability ratio is less than unity for $e\lesssim0.9$, consistent with our initialization at $e_0\leq0.8$. Then as the binary evolves, the decreased semi-major axis increases stability as eccentricity pumping acts to destabilize. The binary reaches a maximum eccentricity of $e\sim0.999$ at $a\sim10^5r_s$, before GWs circularize the binary, but even at this point the stabilization of the decaying binary dominates and we find ${\rm f}^{\rm out}_{\rm st}\lesssim1$. 
Thus, we conclude that the CBD can stably anti-align at the initialization values chosen here, and that it remains stable until at least the point of binary decoupling from the disc, where we end our integration, \ie, we have long-term stable retrograde accretion. This conclusion is in line with previous works, such as \citealt{Nixon2012,Schnittman_Krolik_retro:2015} but in contrast with \citet{RoedigSG_retro:2014} because they considered more massive (by a factor of $\mathcal{O}(10)$) self-gravitating discs.

In the next section, we evolve our systems of interest under the influence of both gas and GWs to find expected binary parameters in the LISA band.

\section{Results}\label{S:Results}

\subsection{Relevant timescales}

We solve the coupled equations in Eq.~\eqref{eq:coupledevo} for the retrograde system for $\fedd=0.1$, $z=1$, an initial total mass $M_{z,0}=10^5~\MSun$, initial spins $s_{1,2}=0$, and for various initial eccentricities $e_0$. In Table~\ref{table:binevo}, we show the corresponding inspiral timescale $t_{\rm ins}$, eccentricity one-year before merger ($e_{1\rm yr}$), the final-to-initial total-mass ratio, and final spins. Here, $M_z/M_{z,0}$ can be well approximated by $\exp(\fedd t_{\rm ins}/50~{\rm Myr})$. The retrograde accretion leads to higher residual eccentricities, which should be measurable in the LISA band \citep{Garg2024a,Garg2024b}. We can also infer from $s_{1,2}>0$ that retrograde accretion leads to component BH spins aligned with their mini-discs and anti-aligned with the binary. Increasing the total mass to $M_{z,0}=10^6~\MSun$ does not meaningfully alter $t_{\rm ins}$ for retrograde binaries,\footnote{Except for $e_0=0$ case, where $t_{\rm ins}\sim250$ Myr.} but $e_{1\rm yr}$ decreases by a factor of 5 from the $M=10^5~\MSun$ case for a given $e_0$. While for a prograde binary, $e_{1\rm yr}\sim 10^{-3}$, irrespective of the initial total mass. Increasing the accretion rate to $\fedd=1$ results in a nearly ten times faster inspiral time as well as slightly higher eccentricities for retrograde CBDs and $e_{1\rm yr} \sim 10^{-2.5}$ for prograde scenarios. We note that previous works have also found trends in eccentricity for prograde vs retrograde consistent with our models \citep{Roedig2011,Schnittman_Krolik_retro:2015,Zrake2021}. While the final spins are almost independent of $\fedd$, we further discuss spin evolution in \S~\ref{Sec:chieff}.
\begin{table}
\centering
    \begin{tabular}{|c|c|c|c|c|}
        \hline
        $e_0$ & $t_{\rm ins}{\rm [Myr]}$&$e_{1\rm yr}$&$M_z/M_{z,0}$&$s_{1,2}$\\
        \hline
        0 & 370 & 0 & 2.03 & 0.99\\
        0.1 & 149 & 0.03 & 1.35 & 0.73\\
        0.2 & 130 & 0.05 & 1.31 & 0.67\\
        0.3 & 113 & 0.06 & 1.26 & 0.62\\
        0.4 & 97 & 0.09 & 1.22 & 0.55\\
        0.5 & 81 & 0.12 & 1.18 & 0.48\\
        0.6 & 64 & 0.16 & 1.14 & 0.41\\
        0.7 & 48 & 0.21 & 1.11 & 0.32\\
        0.8 & 32 & 0.27 & 1.07 & 0.22\\
        \hline
    \end{tabular}
\caption{The initial eccentricity ($e_0$), the inspiral timescale ($t_{\rm ins}$), the eccentricity one year before merger ($e_{1\rm yr}$), the final-to-initial total mass ratio, and final spins assuming $\fedd=0.1$, initial total mass $M_{z,0}=10^5~\MSun$, and zero initial spins, for retrograde evolution.}
\label{table:binevo}
\end{table}

\begin{figure}
\centering
    \includegraphics{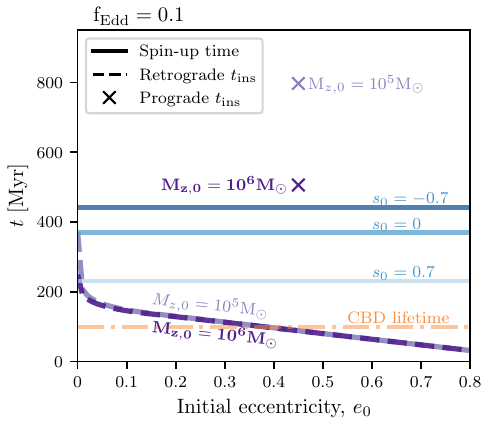}
    \caption{
    Timescales of interest for $\fedd=0.1$ as a function of initial eccentricity $e_0$. We show inspiral timescales (dashed lines) for two initial total masses $M_{z,0}=10^5~\MSun$ (light purple) and $M_{z,0}=10^6~\MSun$ (dark purple) in a retrograde CBD starting from the self-gravitating radius of the disc. For comparison, we show prograde results ($\times$'s) for the same masses initialized at $e_0=0.45$. We also show spin-up timescales (solid lines) for a component BH to increase its spin to $s_i = 0.99$ from three initial spins. We draw an expected, $100$ Myr, CBD lifetime as the dot-dashed orange line.
    }
    \label{fig:timescale}
\end{figure}

In Fig.~\ref{fig:timescale}, as a function of $e_0$, we compare spin-up timescales from different initial spin magnitudes, binary lifetimes for different configurations, and a typical CBD lifetime based on observations of AGN. We see that the prograde systems live longer than their spin-up timescales for any $s_0$, while retrograde systems have lifespans shorter than the spin-up timescale. This implies that prograde systems will always maximally spin up their component BHs to align with the orbital angular momentum \citep{Dotti2010} while in retrograde scenarios this happens only for initially highly spinning component BHs aligned with their minidiscs (and so anti-aligned with the binary, \ie, $s_0\sim1$). We further explore retrograde spin-up in the next sub-section.

The observationally motivated CBD lifetime for $\fedd\gtrsim0.1$ is around $100$ Myr \citep{Hopkins2009}, for masses relevant to LISA. This implies that for $\fedd=0.1$, MBHBs initialized from the self-gravitating radius with either prograde CBDs or retrograde CBDs with low initial binary eccentricity ($e_0\lesssim0.3$) will not reach the LISA band in the absence of other hardening mechanisms. To consider this, \S~\ref{Sec:pop} shows population results for three scenarios: no restriction on inspiral timescale, $t_{\rm ins}\lesssim100$ Myr restriction by starting longer inspiral systems closer than $R_{\rm sg}$, and only considering $R_{\rm sg}$ initialized MBHBs that merge within the CBD lifetime, \ie, $t^{\rm sg}_{\rm ins}<100$ Myr.

\subsection{Spin evolution}\label{Sec:chieff}

Despite retrograde binary lifetimes being shorter than the typical spin-up time, a significant imprint on the effective spin in the LISA band can still arise for retrograde systems.

The effective spin of the binary $\chi_{\rm eff}$ assuming complete (anti-) alignment is \citep{Damour2001}
\begin{align}\label{eq:chieff}
    \chi_{\rm eff}&=\frac{s_{1}+s_{2}}{2}\times\begin{cases} 
        +1, & {\rm for~prograde~CBD}, \\
        -1, &{\rm for~retrograde~CBD}.
        \end{cases}
\end{align}

\begin{figure}
\centering
    \includegraphics{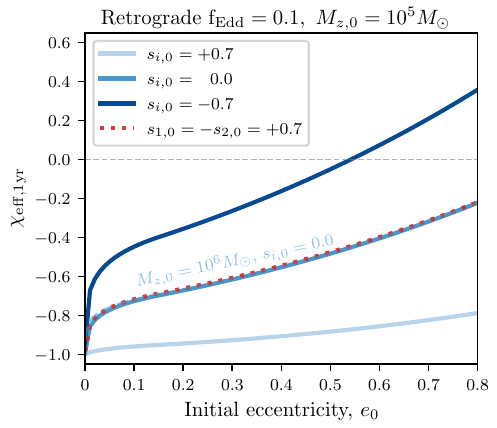}
    \caption{Effective spin $\chiyr$, one year before merger for a $M_{z,0}=10^5~\MSun$ (solid line) MBHB with $\fedd=0.1$ retrograde CBD, as a function of the initial eccentricity $e_0$ for different initial spin configurations. We also include $M_{z,0}=10^6~\MSun$ (dashed medium blue) with zero initial spins for comparison.}
    \label{fig:chieff}
\end{figure}

Fig.~\ref{fig:chieff} plots the expected effective spin one-year before merger ($\chiyr$) for retrograde evolution for a few systems of interests. Low initial eccentricities lead to longer inspiral timescales (see Fig.~\ref{fig:timescale}), allowing more time for the mini-discs to spin-up the BHs, resulting in more negative $\chiyr$ for retrograde binaries. 

\subsection{Population}\label{Sec:pop}

We are interested in estimating the expected distribution for the effective spin ($\chiyr$) and the eccentricity ($e_{\rm 1 yr}$), one-year before merger, when LISA can probe it. To do so we realize a population by drawing the parameters described in Sec.~\ref{S:Model} as follows:
\begin{itemize}
    \item For initial total masses, we relate $M_{z,0}$ to the MBHB's luminosity $L={\rm f}_{\rm Edd}L_{\rm Edd}$, where the Eddington luminosity $L_{\rm Edd}\equiv6.3\times10^4M_{z,0}~{\rm ergs\,g^{-1}\,s^{-1}}$ and then draw $L$ from the observed quasar luminosity function (QLF) at redshifts $z=1$ and $z=2$ \citep{Hopkins2007}. We choose these low redshifts because that's where we expect the highest-SNR MBHB mergers and the most likely candidates for multi-messenger detections \citep[see, \eg][]{Mangiagli2022}. The latter would be particularly helpful in constraining disc properties \citep{Garg2024b}. We are also more likely to detect low eccentricities \citep{Garg2024a} and well-measured effective spins for high-SNR MBHBs as compared to sources at further distances. The former requires SNR~$\gtrsim10$ during inspiral, which is typically $\sim5\%$ of the total event SNR, which is only possible for low-redshift MBHBs. Moreover, to have events in the LISA band, we only consider initial total masses $M_{z,0}\sim[10^{4.5},10^7]~\MSun$ for retrograde MBHBs and $M_{z,0}\sim[10^{4},10^7]~\MSun$ for prograde MBHBs. Different initial mass ranges are needed to get the same distribution of total masses $\sim[10^{4.5},10^{7.5}]~\MSun$ in the LISA band since the prograde inspiral timescale is much longer than in the retrograde case leading to larger increases in mass. 
    \item We draw initial spin magnitudes and eccentricities from uniform distributions: $s_0\sim{\rm Uniform}(-1,1)$ and $e_0\sim{\rm Uniform}(0,0.8)$.
\end{itemize}

\begin{figure}
\centering
\includegraphics{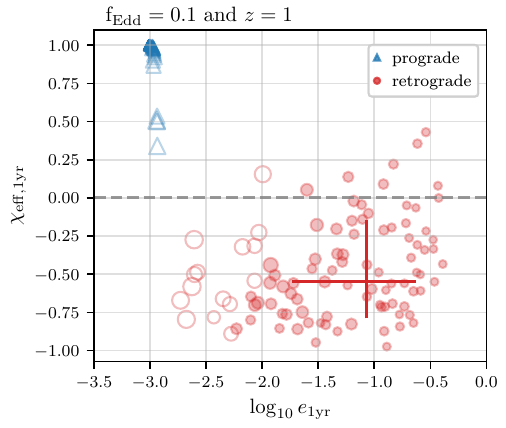}
    \caption{$\chiyr$ vs $e_{\rm 1 yr}$ for 100 LISA-observable MBHBs at $z=1$ for prograde (blue triangles) and retrograde (red circles) CBDs with $\fedd=0.1$ Eddington ratio. The marker size represents how large $M_z$ is in the range $\sim[10^{4.5},10^{7.5}]~\MSun$. Filled symbols corresponds to measurable eccentricity ($e^{\rm det}_{\rm 1 yr}$) as per \citet{Garg2024a,Garg2024b}. For retrograde systems, we draw solid red bars indicating $68$ per cent credible intervals of spin-eccentricity distribution for measurable eccentricity events with intersection indicating median values}
    \label{fig:chieff_ecc_LISA}
\end{figure}

We initialized $100$ MBHBs from their CBD's self-gravitating radius, such that all events have SNR of at least $8$ at $z=1$ in the LISA band to be observable \citep{Garg2022} and to mimic expected MBHB detections by LISA \citep{AmaroSeoane2022}. For both retrograde and prograde MBHBs, we show $\chiyr$ vs $e_{\rm 1 yr}$ in Fig.~\ref{fig:chieff_ecc_LISA} at redshift $z=1$ and ${\rm f}_{\rm Edd}=0.1$. We find that prograde CBDs lead to almost positive unity effective spin and non-measurable $e_{\rm 1 yr}$,\footnote{However, for prograde accretion with $\fedd=1$, we expect measurable eccentricities $e_{1\rm yr}\sim10^{-2.5}$ for $M_z\lesssim10^5~\MSun$.} and in the retrograde case, we mostly have negative $\chiyr$ and measurable eccentricities.\footnote{An equal-mass merging MBHB with $M_z=10^5~\MSun$ has the minimum measurable eccentricity $10^{-2.75}$ at $z=1$ for a one-year observation window \citep{Garg2024b}.}

Assuming all binaries are brought to merger by an accretion phase, Fig.~\ref{fig:chieff_ecc_LISA} shows that one could distinguish between prograde and retrograde from their $\chiyr$ vs $e_{\rm 1 yr}$ correlations in the LISA band, as they occupy different parts of the parameter space. Note that, even for the fastest inspiral case in prograde, we predict almost unity effective spin such that this should also be the case for more complex evolutions $\dot a^{\rm pro}_{\rm gas}$ \citep[see, \eg][]{DOrazio2021}. Similarly, $e_{1\rm yr}$ for prograde scenarios should be the same for equal-mass MBHBs \citep{Zrake2021} as it is only a function of the equilibrium value  $e_{\rm eq}^{\rm pro}=0.45$ \citep[see also][]{Siwek2023, Valli2024}. Thus we expect that our prograde results are robust despite the simplifications. 

\begin{figure}
    \centering
    \includegraphics{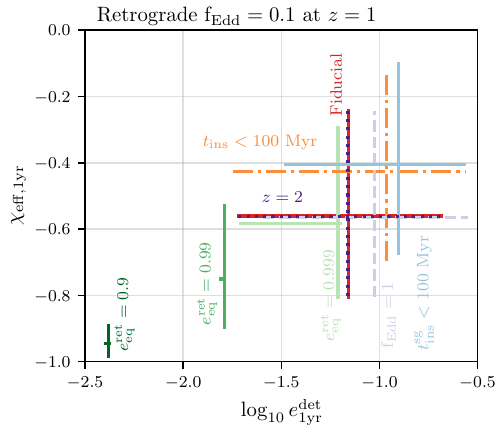}
    \caption{Spin-eccentricity correlations for MBHBs with detectable eccentricity ($e^{\rm det}_{1\rm yr}$) for different scenarios in comparison to our fiducial retrograde CBD (solid red) with $\fedd=0.1$ at $z=1$. We show $\fedd=1$ (dashed light purple), $z=2$ (dotted dark purple), $t_{\rm ins}<100$ Myr (dot-dashed orange), $t^{\rm sg}_{\rm ins}<100$ Myr (solid medium blue), and assuming the maximum eccentricity in retrograde ($e^{\rm ret}_{\rm eq}$) to be either $0.9$ (solid dark green), $0.99$ (solid medium green), and $0.999$ (solid light green). Bars are similar to Fig.~\ref{fig:chieff_ecc_LISA} and we summarize these bounds in Table~\ref{table:chieff_ecc_comp}.
    }
    \label{fig:chieff_ecc_comp}
\end{figure}

In Fig.~\ref{fig:chieff_ecc_comp}, we repeat this experiment for $10^4$ LISA-observable events for varying scenarios to better ascertain the underlying distributions and minimize variance from drawing fewer events. In comparison to our fiducial retrograde case with Eddington ratio $\fedd=0.1$ and redshift $z=1$, we consider $\fedd=1$, $z=2$, and CBD-lifetime restricted inspirals: either $t_{\rm ins}<100$ Myr or only considering events which merge with $t^{\rm sg}_{\rm ins}<100$ Myr, when initialized from the self-gravity radius. We also consider different maximum equilibrium eccentricities (dubbed $e^{\rm ret}_{\rm eq}$) for the retrograde CBD. That is to say, one might imagine that retrograde discs can only pump eccentricity to some maximum value $e^{\rm ret}_{\rm eq}$ at which shock-dissipation at pericenter (or some other mechanism) could shut off eccentricity growth. In such a scenario, the MBHB will stall at $e^{\rm ret}_{\rm eq}$, keep shrinking due to $\dot{a}^{\rm ret}_{\rm gas}$, and will circularize when GWs become dominant. Lowering $e^{\rm ret}_{\rm eq}$ results in longer inspiral timescales and more negative $\chiyr$, but smaller $e^{\rm det}_{1\rm yr}$. We note, however, that each of these scenarios in Fig.~\ref{fig:chieff_ecc_comp} retains an in-band eccentricity above the $e_{1 \rm yr} > 10^{-2.75}$ detection threshold for an equal-mass $10^5 M_\odot$ binary.

\begin{table}
\centering
    \begin{tabular}{|c|c|c|}
        \hline
        Scenario &$\log_{10}e^{\rm det}_{1\rm yr}$&$\chiyr$ \\
        \hline
       Fiducial  & $-1.16_{-0.57}^{+0.48}$ & $-0.56_{-0.25}^{+0.32}$\\
        $\fedd=1$&$-1.03_{-0.57}^{+0.48}$ & $-0.57_{-0.24}^{+0.32}$ \\
        $z=2$ & $-1.16_{-0.56}^{+0.48}$ & $-0.56_{-0.24}^{+0.33}$\\
        $t_{\rm ins}<100$ Myr & $-0.96_{-0.76}^{+0.40}$ & $-0.43_{-0.27}^{+0.30}$\\
        $t^{\rm sg}_{\rm ins}<100$ Myr & $-0.90_{-0.58}^{+0.34}$ & $-0.40_{-0.27}^{+0.31}$ \\
        $e^{\rm ret}_{\rm eq}=0.9$  & $-2.38_{-0.02}^{+0.01}$ & $-0.95_{-0.04}^{+0.06}$\\
        $e^{\rm ret}_{\rm eq}=0.99$  & $-1.79_{-0.03}^{+0.01}$ & $-0.75_{-0.15}^{+0.23}$ \\
        $e^{\rm ret}_{\rm eq}=0.999$ &$-1.21_{-0.50}^{+0.02}$ & $-0.58_{-0.23}^{+0.29}$\\
        \hline
    \end{tabular}
\caption{Measurable eccentricity ($e^{\rm det}_{1\rm yr}$) and effective spin distributions for different scenarios in Fig.~\ref{fig:chieff_ecc_comp}.}
\label{table:chieff_ecc_comp}
\end{table}

Both redshifts produce similar spin-eccentricity correlations, implying that our results are independent of redshift. This is because once we consider detector-frame quantities, a fixed range for total masses $M_{z}\sim[10^{4.5},10^{7.5}]~\MSun$, and the same number of observable events, then only QLF, SNR, and the minimum measurable eccentricity changes with redshift. QLF's shape depends weakly on $z$, SNRs are still higher than our threshold of $8$ at $z=2$, and the minimum measurable eccentricity only increases slightly at $z=2$, effectively, not changing results at different redshifts. While spin distributions are almost the same for both Eddington ratios $\fedd=0.1$ and $\fedd=1$, $e^{\rm det}_{1\rm yr}$ is slightly higher for the latter and this could help to distinguish between Eddington ratios.\footnote{For prograde case, $\fedd=1$ still has unity $\chiyr$ but $e_{1\rm yr}\sim10^{-2.5}$ instead of $e_{1\rm yr}\sim10^{-3}$ for $0.1$ Eddington ratio.} Limiting the inspiral timescale to typical AGN lifetimes of $100$ Myr leads to higher effective spin and higher measurable eccentricities. The former is because smaller initial eccentricities are time limited and not able to spin BHs to high negative effective spin as per Figs~\ref{fig:timescale} and \ref{fig:chieff} and the latter because higher initial eccentricities are selected because they merge more quickly.
%more likely than lower ones due to merging faster. 

\section{Discussion and Conclusion}\label{S:Discussion}

We find that MBHBs with a retrograde CBD typically have negative effective spin and measurable eccentricity in the LISA band.
This is in contrast to prograde scenarios, which result in positive effective spin and smaller or non-measurable $e_{1\rm yr}$. Moreover, the spin-eccentricity distribution of binaries evolving through a retrograde accretion channel occupy a unique part of parameter space that appears to be robust to environmental or evolutionary parameters such as $\fedd$, redshift, or $e^{\rm ret}_{\rm eq}$. However, different parameters could change the exact quantitative spin-eccentricity correlations, which may update with advances in simulating CBD-driven orbital evolution. 

A binary carried to merger via retrograde accretion (with $\fedd\gtrsim0.1$) should have a unique observable signature in the gravitational waveform with a negative effective spin and a detectable eccentricity.
This would likely come in combination with a measurable negative GW-phase perturbation at the $-4$PN order due to secular gas effects in the orbital evolution \citep{Garg2024b}. This would provide a strong clue for distinguishing between binary hardening mechanisms in the LISA band \citep[see, \eg][]{Zwick2023}. 

Chaotic accretion (or similarly many different accretion events leading up to merger, with differing disc alignments) typically leads to low-spin MBHBs near merger \citep{King2008,Chen2023}, which could also be the case for some retrograde MBHBs as shown in Fig.~\ref{fig:chieff_ecc_LISA} but not for the prograde accretion. However, a measurable eccentricity should help in distinguishing retrograde from frequent chaotic accretion, as the latter may not have enough accretion to either pump eccentricities similar to retrograde or sustain equilibrium eccentricity as in prograde. But this needs to be shown via simulations that if a long-term evolution under chaotic accretion can lead to residual measurable $e_{1\rm yr}$. 
Also, measurable $-4$PN-order GW-phase shifts should help to differentiate retrograde accretion from hardening via a third MBH \citep{Bonetti2019}, which can also excite high residual eccentricities in the LISA band. Moreover, a CBD leads to an aligned spinning binary while hardening via a third MBH typically forms precessing MBHBs near-merger. Additionally, if we detect an EM counterpart for an MBHB that confirms the presence of an accretion disc \citep[see, \eg][]{Haiman2009}, then the measured effective spin can provide evidence for the disc orientation. 

We have relied on a number of assumptions that might be improved upon in the future. While the hydrodynamical simulation results in \citet{Tiede2024b} are only computed for eccentricity up to $0.8$, we extrapolate them to higher eccentricities. Furthermore, existing hydrodynamical results that span a sufficient range of orbital eccentricities derive from 2D, isothermal hydrodynamical treatments of the CBD. While the key ingredients of retrograde eccentricity pumping \citep[for, \eg, 3D, self-gravitating discs, including magnetic feilds, or unequal masses][]{NixonKing_retro+2011, RoedigSG_retro:2014, BankertShiKrolik:2015, AmaroSeoane:RetroDiscs:2016} and retrograde mini-discs \citep{Tiede2024b} appear to be robust, inclusion of more complete physics -- \eg~radiative transfer, general relativistic disc-response -- in future simulations will be necessary to improve the accuracy of the predictions presented here. 
Additionally, we add evolution rates from gas and GWs linearly as its not possible to infer cross terms from current simulations for MBHBs. However, studies by \citet{Derdzinski2019,Derdzinski2021} suggest that for extreme mass ratio inspirals cross terms should be weak, which may also extend to MBHBs. Even though we checked that our systems are never at small enough binary separation to require relativistic corrections, \citet{Peters1964} may still breakdown for extremely high eccentricities due to apsidal precession inducing small but possibly non-negligible error. However, this is not clear how it will affect our results and we leave this for future work. We additionally only consider equal-mass MBHBs, due to the limitation of retrograde results and under the assumption that before reaching the self-gravitating radius the binary has become close to equal-mass through preferential accretion. Lastly, typical hierarchical clustering models suggest that most MBHB mergers may happen at redshifts larger than $z=2$ \citep[see, \eg][]{Barausse2012,Bonetti2019}, implying that the number of events may not be enough to do population inference.

In summary, we have demonstrated a possibly unique signature of the retrograde CBD accretion channel comprising MBHB mergers with negative effective spin and measurable eccentricities. This can help to distinguish between the various scenarios---accretion, stellar hardening, and third-body interactions---that enable massive binaries to merge within the lifetime of the universe.

\section*{Data availability statement}

The data underlying this article will be shared on reasonable request to the authors.

% =============================================================================
\section*{Acknowledgements}
MG acknowledge support from the Swiss National Science Foundation (SNSF) under the grant 200020\_192092. D.J.D. and C.T. received support from the Danish Independent Research Fund through Sapere Aude Starting grant No. 121587.  C.T. also received support from European Union’s Horizon 2023 research and innovation program under Marie Sklodowska-Curie grant agreement No. 101148364. We thank the anonymous referee for helpful comments that improved this work. The authors also acknowledge use of the NumPy \citep{harris2020array} and Matplotlib \citep{Hunter2007}.
% =============================================================================
\scalefont{0.94}
\setlength{\bibhang}{1.6em}
\setlength\labelwidth{0.0em}
\bibliographystyle{mnras}
\bibliography{retrobhb}
\normalsize

\appendix
\input{Appendices/Stability}

\bsp %typesetting comment
\label{lastpage}
\end{document}

%% file: Appendices/Stability.tex
\section{Stability of retrograde accretion}\label{App:fst}

The accretion stability ratio, \ie, $L_{\rm CBD}=2\pi\int_0^{Na} r {\rm d}r \Sigma(r)\sqrt{GM_z r}  $ divided by $2L_{\rm Binary}=2M_z\sqrt{GM_za(1-e^2)}/4$, is
\begin{equation}\label{eq:stability2}
    {\rm f}_{\rm st}=\frac{L_{\rm CBD}}{2L_{\rm Binary}}=4\pi\frac{\int_0^{Na}{\rm d}r\Sigma(r) r^{\frac32}}{M_z\sqrt{a(1-e^2)}}.
\end{equation}

Depending on the radial coordinate, we can divide the \citet{ShakuraSunyaev1973} disc into three regions - inner, middle, and outer - with different respective surface density profiles \citep{Haiman2009,Haiman2022_Err}. Substituting into Eq.~\eqref{eq:stability2} gives

\begin{align}
    {\rm f}^{\rm in}_{\rm st}(a)&={\rm f^{(I)}}_{\rm st}(a),~a\lesssim N_1r_s,\\
    {\rm f}^{\rm mid}_{\rm st}(a)&={\rm f^{(I)}}_{\rm st}(N_1r_s)+{\rm f^{(II)}}_{\rm st}(a),~N_1r_s\lesssim a\lesssim N_2r_s,\nonumber\\
    {\rm f}^{\rm out}_{\rm st}(a)&={\rm f^{(I)}}_{\rm st}(N_1r_s)+{\rm f^{(II)}}_{\rm st}(N_2r_s)+{\rm f^{(III)}}_{\rm st}(a),~N_2r_s\lesssim a\leq Na,\nonumber
\end{align}
where  $N_1\ll299\left({\fedd}/{0.1}\right)^{16/21}\left({M_z}/{10^5\MSun}\right)^{2/21}$, $N_2\sim4100\left({\fedd}/{0.1}\right)^{2/3}$, and for $\epsilon\sim0.1$:\footnote{We ignore the lower integration limit as it is much smaller than the upper limit for simplicity.}
\begin{align}\label{eq:stability3}
    {\rm f^{(I)}}_{\rm st}(a)\sim&\frac{1.85\times10^{-6}}{\sqrt{1-e^2}}\left(\frac{\alpha}{0.1}\right)^{-1}\left(\frac{\fedd}{0.1}\right)^{-1}\left(\frac{M_z}{10^5\MSun}\right)^{-\frac{5}{2}}\\
    &\left(\frac{a}{500r_s}\right)^{\frac72},\nonumber\\
    {\rm f^{(II)}}_{\rm st}(a)\sim& \frac{1.46\times10^{-5}}{\sqrt{1-e^2}}\left(\frac{\alpha}{0.1}\right)^{-\frac45}\left(\frac{\fedd}{0.1}\right)^{\frac35}\left(\frac{M_z}{10^5\MSun}\right)^{-\frac{1}{5}}\nonumber\\
    &\left(\frac{a}{5000r_s}\right)^{\frac75},\nonumber\\
    {\rm f^{(III)}}_{\rm st}(a)\sim &\frac{0.39}{\sqrt{1-e^2}}\left(\frac{\alpha}{0.1}\right)^{-\frac45}\left(\frac{\fedd}{0.1}\right)^{\frac{7}{10}}\left(\frac{M_z}{10^5\MSun}\right)^{-\frac{1}{20}}\left(\frac{N}{3}\right)^{\frac74}\nonumber\\
    &\left(\frac{a}{5.32\times10^6r_s}\right)^{\frac54},\nonumber
\end{align}
where we have normalized $a$ by a multiplier of $r_s$ larger than respective regions of validity up to the self-gravitating radius for the outer disc. Given that ${\rm f^{(I)}}_{\rm st}\ll{\rm f^{(II)}}_{\rm st}\lll{\rm f^{(III)}}_{\rm st}$, we have
\begin{align}
    {\rm f}^{\rm in}_{\rm st}\sim&{\rm f^{(I)}}_{\rm st}(a),\\
    {\rm f}^{\rm mid}_{\rm st}\sim&{\rm f^{(II)}}_{\rm st}(a),\nonumber\\
    {\rm f}^{\rm out}_{\rm st}\sim&{\rm f^{(III)}}_{\rm st}(a).\nonumber
\end{align}

All three regions are stable for our parameters of interest.